\documentclass[pre,aps,10pt,twocolumn]{revtex4-1}
\usepackage{geometry}
\usepackage{graphicx}
\usepackage{amsfonts}
\usepackage{amsmath}
\usepackage{esint}
\usepackage{setspace}
\usepackage{bm}
\usepackage{url}
\usepackage{color}
\usepackage{xfrac}
\usepackage{array,url}

\newcommand{\av}{{\bm a}}
\newcommand{\bv}{{\bm b}}

\newcommand{\dv}{{\bm d}}
\newcommand{\Rv}{{\bm R}}

\newcommand{\fv}{{\bm f}}
\newcommand{\ev}{{\bm e}}
\newcommand{\tv}{{\bm t}}

\newcommand{\rv}{{\bm r}}

\newcommand{\qv}{{\bm q}}

\newcommand{\zv}{{\bm z}}

\newcommand{\Gv}{{\bm G}}

\newcommand{\kb}{k}  
\newcommand{\Wr}{{\rm W}}

\begin{document}
\title{Mechanical Graphene}

\author{Joshua E.~S.~Socolar}
\affiliation{Department of Physics, Duke University, Durham, NC}
\author{Tom C.~Lubensky}
\affiliation{Department of Physics and Astronomy, University of Pennsylvania,
Philadelphia, PA}
\author{Charles L.~Kane}
\affiliation{Department of Physics and Astronomy, University of Pennsylvania,
Philadelphia, PA}

\date{\today}

\begin{abstract}
We present a model of a mechanical system with a vibrational mode spectrum identical to the spectrum of electronic excitations in a tight-binding model of graphene.  The model consists of point masses connected by elastic couplings, called ``tri-bonds,'' that implement certain three-body interactions, which can be tuned by varying parameters  that correspond to the relative hopping amplitudes on the different bond directions in graphene.  In the mechanical model, this is accomplished by varying the location of a pivot point that determines the allowed rigid rotations of a single tri-bond.  The infinite system constitutes a Maxwell lattice, with the number of degrees of freedom equal to the number of constraints imposed by the tri-bonds.  We construct the equilibrium and compatibility matrices and analyze the model's phase diagram, which includes spectra with Weyl points for some placements of the pivot and topologically polarized phases for others.   We then discuss the edge modes and associated states of self stress for strips cut from the periodic lattice.    Finally, we suggest a physical realization of the tri-bond, which allows access to parameter regimes not available to experiments on (strained) graphene and may be used to create other two-dimensional mechanical metamaterials with different spectral features.
\end{abstract}

\maketitle

\section{Introduction}

Topology \cite{Nakahara2003,volovik03,Volovik2007} has become
an important tool in advancing our understanding of electronic
properties of solids. It plays an important role, for example,
in determining the nature of surface states in a wide variety
of systems including polyacetylene \cite{ssh,jackiw76}, quantum
Hall systems \cite{halperin82,haldane88}, and topological
insulators
\cite{km05b,bhz06,mb07,fkm07,HasanKane2010,QiZhang2011}.  The
success in applying topological ideas to electronic systems has
recently inspired their generalization to certain classes of
classical mechanical systems
\cite{KaneLub2014,LubenskySun2015,PauloseVit2015,PauloseVit2015-b,ChenVit2014,VitelliChe2014,Chen2015,Xiao2015,Po2014,Yang2015,Nash2015,Wang2015,Wang2015a,Susstrunk2015,Kariyado2015,Peano2015,Mousavi2015,Khanikaev2015,RocklinLub2016}
by establishing a correspondence between quantum electronic
Hamiltonians and the ``square root" of the mechanical dynamical
matrix. 
This correspondence is exact for a class of Maxwell lattices for which
there is a balance between the number of degrees of freedom and the number
of constraints per unit cell.  A prototype example is the one-dimensional
mechanical model~\cite{KaneLub2014,ChenVit2014,ZhouVitelli2016} whose excitation
spectrum precisely matches the spectrum of the
Su-Schrieffer-Heeger (SSH) model for polyacetylene. In the
one-dimensional SSH model, electrons move on a bipartite
lattice with different hopping matrix elements on alternating
bonds that connect the sites on the $A$- and $B$-sublattice.  The
mechanical model consists of rigid rotors with pivot points fixed on the
$A$-sublattice and connected by central-force springs residing
on the the $B$-sublattice.  There is a one-to-one
correspondence between electrons on the $A$-sublattice in the
SSH model and the rotors in mechanical model and between
electrons on the $B$-sublattice and the springs of the
mechanical model.
We note that in contrast to other models of topological mechanics~\cite{Yang2015,Nash2015,Wang2015,Wang2015a,Susstrunk2015,Kariyado2015,Peano2015,Mousavi2015,Khanikaev2015} Maxwell lattices exhibit topologically protected zero frequency modes, along with an intrinsic particle-hole symmetry in the analog quantum Hamiltonian.

\begin{figure}
\includegraphics[width=1.0\columnwidth]{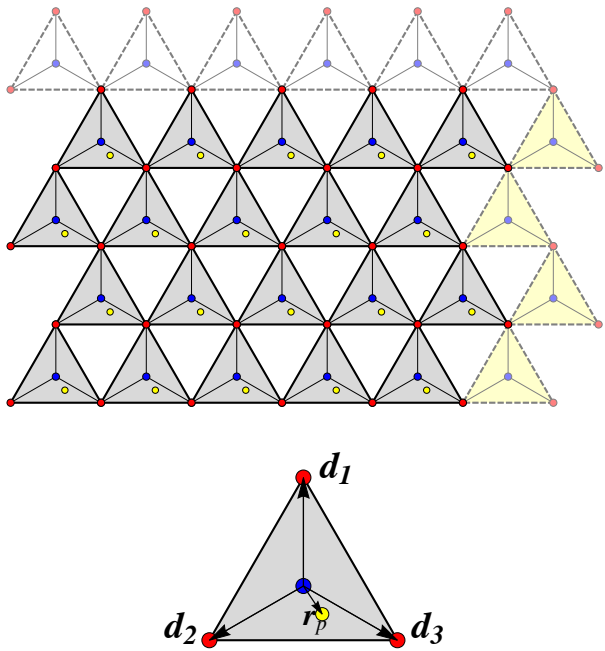}
\caption{Top:  The tri-bond lattice.  Each shaded triangle represents a tri-bond
centered on a $B$ site of the graphene lattice (blue disks).  The tri-bonds
are connected at their vertices, which lie on $A$-sublattice (red disks).
The yellow disk in each tri-bond marks the pivot point.  Dashed tri-bonds
are those that are cut to create free surfaces.  Bottom: A single tri-bond.
The pivot point $\rv_p$ is at $(x_1\dv_1+x_2\dv_2+x_3\dv_3)$.}
\label{fig:model}
\end{figure}

Here, we introduce and explore the properties of a model mechanical system whose bulk vibrational excitations are in correspondence with the electronic excitations of the particle-hole symmetric two-band tight-binding model of graphene. 
 We consider a generalized
graphene model  describing nearest neighbor hopping on
a honeycomb lattice in which the hopping matrix elements
for the three bonds emanating from a given site are in general
different.   This model can represent strained graphene \cite{KaneMele1997} for 
modest variations in the hopping amplitudes, and also 
arises in the analysis of Kitaev's honeycomb lattice model \cite{Kitaev2006}. Our
mechanical analog  is
constructed by closely following the paradigm used in the
construction of the SSH analog.  The honeycomb lattice of
graphene, like the 1D SSH lattice, is bipartite with $A$- and
$B$-sublattices, shown as red and blue disks in Fig. \ref{fig:model}. Since there
is only one degree of freedom per site in the graphene model,
we assign a scalar variable $z(\Rv_A)$, which we identify as
vertical height displacement, to each site on the
$A$-sublattice, and we assign a kinetic energy
$\dot{z}^2(\Rv_A)/2$ to that site.  On each site in the
$B$-sublattice, we need to assign the analog of a bond
connecting sites on the $A$-sublattice, but this ``bond" is
connected to three rather than the usual two sites. To each of
these triangular bonds, which we call \emph{tri-bonds}, we
assign a kind of ``stretching" energy that depends
quadratically on a certain linear combination of the heights at
the three sites.
Figure \ref{fig:model} shows our
model mechanical graphene lattice, with red sites representing
the locations of the height variables and gray triangles representing
the tri-bonds.

The tri-bonds are the analogs of springs in a standard
ball-and-spring lattice, and each imposes a constraint that
leads to a generalization of the Maxwell-Calladine theorem
\cite{Calladine1978} relating the numbers of zero modes and
states of self stress (SSS) to the number of degrees of freedom
and number of constraints.
Under periodic boundary conditions, our mechanical graphene
lattice is a generalized Maxwell lattice
\cite{LubenskySun2015,Calladine1978} in which the number of
constraints equals the number of degrees of freedom, and it
exhibits all of the properties of ball-and-spring Maxwell
lattices: (1) Each zero mode in the bulk spectrum, which again
matches the electronic spectrum of graphene, is accompanied by
a SSS. (2) Each lattice is characterized by a topological
polarization or by Weyl modes. (3)  Periodic strips or finite
lattices cut from a periodic lattice whose spectrum is fully
gapped have a number of zero-frequency surface modes equal to
the number of tri-bonds cut, i.e., at least one zero mode per
surface wavenumber, residing on one or the other of the
opposite free edges. (4) The number of zero modes at a given
wavenumber on a free surface is determined by the topological
polarization or by the positions of Weyl modes and by a local
surface polarization. (5) Domain walls connecting lattices of
different topological polarization harbor either zero modes or
self stress for each wavenumber along the wall; those
connecting different Weyl lattices have zero modes or states of
self stress for some wavenumbers but not for others. 

Though
the bulk spectrum of our model and graphene are identical,
their surfaces modes are not.   The top and bottom edges of a horizontal strip of
the mechanical lattice, which can be created by removing the
row of horizontal of dashed tri-bonds in Figure \ref{fig:model}
from a periodic lattice, are different: the top surface exposes
tri-bond vertices and the bottom a straight continuous line of
tri-bond edges.  These two surfaces correspond, respectively,
to the bearded (with dangling bonds) and zigzag edges of
graphene~\cite{Kohmoto2007}.  It is not possible to create a strip in the
mechanical lattice like that in graphene in which both edges
are zigzagged. Both edges of vertical strips exhibit a
two-tri-bond zigzag pattern.  The corresponding graphene edges
correspond to armchair edges with an extra dangling bond at
every second row.

This paper is divided into five sections of which this is the
first.  Section~\ref{sec:mech-graphene} presents details of our
model and defines the equilibrium and compatibility matrices
that establish the Maxwell-Calladine theorem.
Section~\ref{sec:phasediagram} treats the excitation spectrum
and establishes a phase diagram with a region harboring Weyl
points and regions that carry a topological polarization.
Section~\ref{sec:edge} discusses zero modes at free edges or in
domain walls. Section~\ref{sec:phys-models} presents physical
models for the tri-bonds.

\section{Mechanical Graphene Model\label{sec:mech-graphene}}

Figure~\ref{fig:model} provides a visual image of our model. We
take the Bravais lattice constant to be $a$ and define
primitive reciprocal lattice vectors
\begin{eqnarray}
{\bf a}_1 &=& a \hat x ,\\
{\bf a}_2 &=& a (-{1\over 2}\hat x + {\sqrt{3}\over 2}\hat y), \\
{\bf a}_3 &=& a (-{1\over 2}\hat x - {\sqrt{3}\over 2}\hat y).
\end{eqnarray}
We also define the three vectors from a tri-bond centroid to
its vertices:
\begin{eqnarray}
{\bf d}_1 &=& {a\over \sqrt{3}} \hat y \\
{\bf d}_2 &=& {a\over\sqrt{3}}(-{\sqrt{3}\over 2}\hat x - {1\over 2}\hat y) \\
{\bf d}_3 &=& {a\over\sqrt{3}}({\sqrt{3}\over 2}\hat x - {1\over 2}\hat y).
\end{eqnarray}

The centroids of the tri-bonds are located on $B$ sites of the
honeycomb and their vertices lie at $A$ sites. Each site on the
$A$-sublattice is occupied by a unit mass that is constrained
(say by a frictionless rod) to move in the vertical direction.
We assume these vertical displacements $z(\Rv_A)$ are small
enough that linear approximations can be used.  Each
tri-bond is pinned at a pivot point that is displaced from its
centroid by
\begin{equation}
\rv_p = x_1 \dv_1 + x_2 \dv_2 + x_3 \dv_3 ,
\end{equation}
and the three tri-bonds meeting at a given site are connected
in a way that allows each to freely rotate about an axis
passing through its pivot point and the site in question.

Any location of the pivot point (within or outside the triangle
spanned by the tri-bond vertices) can be specified by a unique
triple $x=(x_1,x_2,x_3)$ with $x_1+x_2+x_3 = 1$. With this
parametrization, the condition
\begin{equation}
e(\Rv_B) \equiv  \sum_{i=1}^3 x_i z(\Rv_B + \dv_i)= 0,
\label{eq:stretch}
\end{equation}
is satisfied for any rigid rotation of the tri-bond, where
$\Rv_B$ is the $B$-sublattice location of the tri-bond centroid
and $\Rv_B + \dv_i$ are the $A$-sublattice positions of the
vertices of the tri-bond. Violations of this condition
necessarily cause a distortion of the tri-bond that costs some
energy,
which to lowest order in displacements must be
quadratic in $e(\Rv_B)$, giving us an elastic energy
\begin{equation}
U= \frac{1}{2} \kb \sum_B e^2 (\Rv_B)
\label{eq:pot_E}
\end{equation}
where the sum is over all sites $\Rv_B$ in the $B$-sublattice.
This energy has exactly the same form as that of a lattice of
harmonic springs where the sum is over bonds and $e$ is the
elongation of a bond.  Following this analogy, we call
$e(\Rv_B)$ the \emph{stretch} of the tri-bond at $\Rv_B$. Of
course, with the aid of Eq.~(\ref{eq:stretch}) we can express
$U$ as a function of the displacements $z(\Rv_A)$ instead of
stretches $e(\Rv_B)$.

Given $U$, we can now construct expressions for the $N_B$
tri-bond \emph{tensions} $t(\Rv_B)$ conjugate to stretches of
the tri-bonds and the $N$ site forces $f(\Rv_A)$ conjugate to
vertical displacements of the masses:
\begin{eqnarray}
t(\Rv_B) &  = & \frac{\partial U}{\partial e(\Rv_B)} = \kb e(\Rv_B) \nonumber \\
& = & \kb \sum_i x_i z(\Rv_A + \dv_i) \label{eq:tv}\\
f(\Rv_A) & = & \frac{\partial U}{\partial z(\Rv_A)} =  \kb \sum_{i} x_i e(\Rv_A - \dv_i) \nonumber \\
& = & \sum_i x_i t(\Rv_A-\dv_i)  .
\label{eq:fv}
\end{eqnarray}
The tension $t$ is one that induces ``stretching'' of the
tri-bond in a manner exactly analogous to the tension in a
spring bond inducing stretching of the spring. The tri-bond
stretch is a measure of the deviation from coplanarity of its
three vertices and pivot point.

Introducing the $N$-dimensional vectors $\zv$ and $\fv$ of
site-displacements and forces and the $N_B$ dimensional vectors
of $\ev$ and $\tv$ of stretches and tensions, we can write
Eqs.~(\ref{eq:stretch}), (\ref{eq:tv}), and (\ref{eq:fv}) as
\begin{equation}
\ev = C \zv; \qquad \tv = \kb \ev; \qquad \fv= Q \tv,
\end{equation}
where $C$ is the compatibility matrix with components
\begin{equation}
C(\Rv_B, \Rv_A) = \sum_i x_i \delta_{\Rv_B + \dv_i, \Rv_A} ,
\end{equation}
and $Q$ is the equilibrium matrix with components
\begin{equation}
Q( \Rv_A, \Rv_B) = \sum_i x_i \delta_{\Rv_A - \dv_i, \Rv_B} .
\end{equation}
$Q = C^T$ as required. In Fourier space,
\begin{equation}
e(\qv) = C(\qv) z(\qv); \qquad f(\qv) = Q(\qv) t(\qv) ,
\end{equation}
where $C(\qv)$ and $Q(\qv)$ in this case are $1 \times 1$
matrices with
\begin{equation}
C(\qv) = \sum_i x_i e^{i \qv\cdot \dv_i} = Q^*(\qv) .
\end{equation}

As in the case of central-force springs, the null space of $C$
consists of zero modes and that of $Q$ of SSSs. The global and
wave-number specific Maxwell-Calladine index theorems~\cite{LubenskySun2015} follow
immediately:
\begin{equation}
N_0 - S= N-N_B
\end{equation}
and
\begin{equation}
n_0(\qv) - s(\qv) = n-n_B ,
\end{equation}
where $N_0$ and $S$ are, respectively, the total number of zero
modes and the total number of SSSs, $n_0(\qv)$ and $s(\qv)$ are
the numbers of zero modes and states of self stress at
wavenumber $\qv$, and $n$ and $n_B$ are the number of sites (1
under PBC) and number of tri-bonds (1 under PBC) per unit cell.

The energy can be written in various ways in terms of these
variables:
\begin{eqnarray}
E & = &\frac{k}{2 N} \sum_{\qv} |e(\qv)|^2 \\
& = & \frac{1}{2 N}\sum_{\qv} z(-\qv) D(\qv) z(\qv) \\
& = & \frac{1}{2 k N} \sum_{\qv} |t(\qv)|^2 ,
\end{eqnarray}
where $D(\qv) = k|C(\qv)|^2$ is the (one-dimensional) dynamical
matrix. For the system with PBCs, the corresponding quantum
Hamiltonian is block diagonal with $2\times 2$ blocks of the
form
\begin{equation}
H(\qv) = \omega_0 \left(\begin{array}{cc} 0 & C^* (\qv) \\C(\qv) & 0 \end{array}\right),
\end{equation}
where we define the normal mode frequency scale $\omega_0 = \sqrt{k}$.
The square of this Hamiltonian is diagonal with lower entry $k
C C^*$ and upper entry $k C^* C =D$ (which are identical as $C$
is one-dimensional). The eigenvalues of $H(\qv)$, given by $\pm \omega(\qv)$, are simply the two
square roots of $D$ and thus specify the normal mode dispersion $\omega(\qv)$, with $C(0)=1$ implying
$\omega_0 = \omega(\qv=0)$.  

If we identify the bond hopping amplitudes $t_i = \omega_0 x_i$, then $H(\qv)$ is formally identical to the Hamiltonian of the nearest neighbor tight binding model of graphene~\cite{KaneMele1997}.   Equivalently, the constraint $\sum x_i = 1$ corresponds to setting $\omega_0 = \sum_i t_i$.

\section{Weyl Points and Phase Diagram
\label{sec:phasediagram}}

The spectrum and topological properties of mechanical graphene
depend on the location of the pivot point in the tri-bond. The
features of the different models can thus be represented on a
ternary phase diagram in which each point $(x_1,x_2,x_3)$
corresponds precisely to the placement of the pivot point.
Unlike a typical ternary phase diagram, however, negative
values of $x_i$ (corresponding to pivot points outside the triangle) are allowed here.

\begin{figure}
\includegraphics[width=\columnwidth]{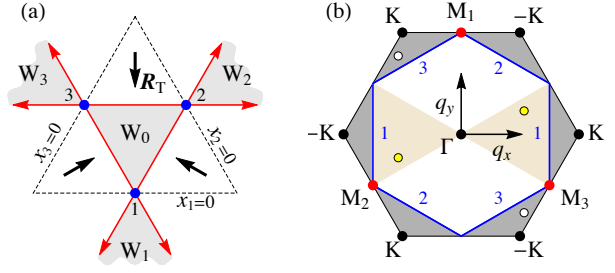}
\caption{(a) Ternary phase diagram of strained graphene.  The dashed triangle marks the boundary of the region in which all $x_i$'s are positive.  In each region marked $\Wr_i$, there are two Weyl points in the Brillouin zone.  In the other regions, the bands are characterized by nonzero winding numbers when traversed in the directions indicated by arrows.  (b)  The Brillouin zone with high symmetry points marked, including ${\bf K} = (4\pi/3a,0)$ (and equivalent points)
and ${\bf M}_1 = (0,2\pi/\sqrt{3}a)$.  Dark gray (light tan) shaded regions indicate possible locations of Weyl points corresponding to region $\Wr_0$ ($\Wr_1)$, with white (yellow) disks showing generic possible arrangements.
At each blue point  $i$ ($=1,2,3$) in (a), there are zero modes along the pair of blue lines marked $i$ in (b).
Along each red line in (a), there is a red point in (b) containing degenerate Weyl points.}
\label{fig:phasediagram}
\end{figure}

In this section, we will derive the phase diagram shown in
Fig.~\ref{fig:phasediagram}(a), in which there are three gapped
regions with different topological polarizations $\Rv_T$ and
four regions, $\Wr_{0,1,2,3}$, with Weyl points.  The point $x=(1/3,1/3,1/3)$
corresponds to undistorted graphene.  Other points correspond
to strained graphene, though the regions outside $\Wr_0$
correspond to degrees of strain that are probably too large to be
physically realizable.

\begin{figure}
\includegraphics[width=1.0\columnwidth]{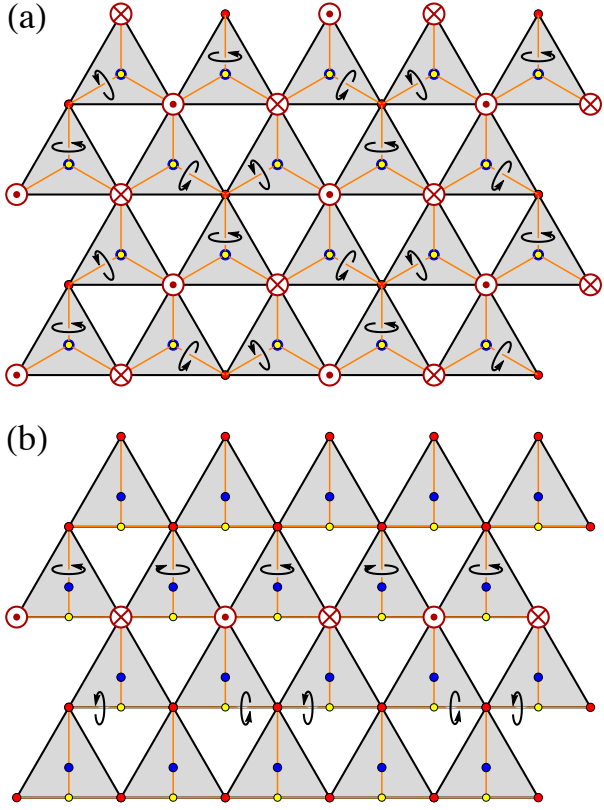}
\caption{(a) A Weyl mode for $x=(1/3,1/3,1/3)$, corresponding to unstrained graphene.  The pivot is at the centroid of the tri-bond.  (b) A zero mode for $x=(0,1/2,1/2)$.  The mode is confined to a single line.  The pivot point is at the midpoint of the bottom edge, and each tri-bond rotates rigidly about its pivot so that vertices move in and out of the plane as indicated.}
\label{fig:modes}
\end{figure}

Weyl points arise when $C({\qv}) = 0$, which generically occurs
for pairs of points $\pm {\qv}^*$, where ${\qv}^*$ is a
solution of
\begin{equation}\label{eqn:weyl}
C(\qv^*) = x_1 e^{i {\qv}^* \cdot {\dv}_1} + x_2 e^{i{\qv}^*\cdot {\dv}_2} + x_3e^{i{\qv}^*\cdot {\dv}_3} = 0.
\end{equation}
When
$x=(1/3,1/3,1/3)$ (at the middle of $\Wr_0$ in
Fig.~\ref{fig:phasediagram}(a)) Eq.~(\ref{eqn:weyl}) is satisfied
at the $\pm {\rm K}$ points $\qv^* = \pm \qv_{\rm K}$, where
\begin{equation}
\qv_{\rm K} = {4\pi\over {3a}}{\hat x} \quad\Rightarrow\quad
e^{i \qv_{\rm K} \cdot {\dv}_j} = e^{2\pi i (j-1) /3}.
\end{equation}
 This corresponds to the well known Weyl point at the Brillouin zone corner in unstrained graphene.
In Fig.~\ref{fig:modes}(a) we show the displacements of one of the zero frequency modes at $\qv_{\rm K}$.   It is straightforward
to see in Fig.~\ref{fig:modes}(a) that this is indeed a ``floppy mode'' in which, to linear order, each tri-bond undergoes a rigid rotation about its pivot point and hence is not stretched.

Because the phase of $C(\qv)$ advances by $2\pi$ as $\qv$ wraps
around $\qv^*$, the Weyl points are locally protected and cannot
be removed by a smooth deformation.   Therefore, there must be a Weyl phase
in a finite region around $(1/3,1/3,1/3)$.   However, it is also clear from (\ref{eqn:weyl}) that
for $x = (1,0,0)$, $(0,1,0)$ or $(0,0,1)$ there are no solutions to
$C(\qv^*)=0$.  Thus, there must be a gapped phase in the
vicinity of the three corners of the dashed triangle in Fig.~\ref{fig:phasediagram}(a).

To determine the phase boundaries, we note that Weyl points can only
only disappear if they meet, which must occur at a
time-reversal-invariant point ${\qv}^* =-{\qv}^* + {\Gv}$,
where $\Gv$ is a reciprocal lattice vector. ${\qv}^*=0$ is
ruled out for finite $x$ because $C(0) = x_1+x_2+x_3 = 1$, so
this must occur at one of the three M points
\begin{equation}
\qv_{{\rm M}j} = {2\pi\over {\sqrt{3}a}} \hat{\dv}_j,
\end{equation}
which satisfy
\begin{equation}
\qv_{{\rm M}j} \cdot {\dv}_k = \left\{\begin{array}{ll}  2\pi/3 & j=k \\
-\pi/3 & j\ne k \end{array}\right. .
\end{equation}
Then Eq.~(\ref{eqn:weyl}) requires
\begin{equation}
x_k - x_{k-1} - x_{k+1} = 0
\label{eqn:boundary}
\end{equation}
where the subscript is defined cyclically.  Together with $\sum
x_i = 1$, this implies $x_k = 1/2$ and $x_{k-1}+x_{k+1} = 1/2$,
which define the three red lines bounding the triangle inscribed in
the dashed region of 
Fig.~\ref{fig:phasediagram}(a).  The Weyl phase in the vicinity of 
$x=(1/3,1/3,1/3)$ thus corresponds to the region $\Wr_0$.   On the boundary of
$\Wr_0$ the Weyl points meet and annihilate.

The points outside the dashed triangle of Fig.~\ref{fig:phasediagram}(a)
have the pivot
outside the tri-bond and have one or two of the $x_i$ negative.  This corresponds
in the graphene model to having bond(s) with a negative hopping amplitude.
Systems with negative hopping amplitudes are closely related to systems with positive
hopping amplitudes.   The sign of the hopping amplitude on one of the three bonds
(say along ${\bf d}_1$) can be changed by a non-uniform gauge transformation that changes the signs of all sites on every other horizontal (zig-zag) line of bonds on the honeycomb lattice. This transformation takes bond hopping amplitudes $(t_1,t_2,t_3) \rightarrow (-t_1,t_2,t_3)$, which in our mechanical model takes $x =(x_1,x_2,x_3) \rightarrow x'=(-x_1,x_2,x_3)/(x_2+x_3-x_1)$.   This gauge transformation does not change the normal mode frequencies except for an overall constant factor due to the relation between $x_i$ and $t_i$.   However, since the gauge transformation is at a nonzero wavevector $\qv_{{\rm M}1}$, it leads to a shift in the wavevector of the normal modes and hence a transformation of the dispersion relation:
\begin{equation}
\omega_{x'}(\qv) = (x_2+x_3-x_1)^{-1} \omega_{x}(\qv + \qv_{{\rm M}1}).
\label{gaugetransform}
\end{equation}
This transformation maps $x = (1/3,1/3,1/3)$ to $x' = (-1,1,1)$, so the normal mode spectra at these two pivot point locations are related by Eq.~(\ref{gaugetransform}).   In particular, for $x'$ there are Weyl points
at $\qv_{\rm K} + \qv_{{\rm M}1}$.   
Indeed, Eq.~(\ref{eqn:weyl}) is satisfied for $\qv^* = \pm \qv_{\rm L}$, where
\begin{equation}
\qv_{\rm L} ={2\pi\over {3a}}{\hat x}
\quad\Rightarrow\quad
e^{i \qv_{\rm L} \cdot {\dv}_{(1,2,3)}} =  (e^0,e^{\pi i /3},e^{-\pi i/3}).
\end{equation}
It can be checked that $\qv_{\rm L} = - \qv_{\rm K} + \qv_{{\rm M}1}$ up to a reciprocal lattice vector.
More generally, the Weyl phase $\Wr_0$ maps to the region $\Wr_1$ in Fig.~\ref{fig:phasediagram}(a).
Similar transformations identify the Weyl phases $\Wr_2$ and $\Wr_3$, whose boundaries are again
given by the red lines defined by Eq.~(\ref{eqn:boundary}).   

When $x$ is in $\Wr_0$ in Fig.~\ref{fig:phasediagram}(a),
the Weyl points are at ${\pm\qv}^*$, which reside in the dark
gray regions of Fig.~\ref{fig:phasediagram}(b).
When $x$ is in $\Wr_1$, the Weyl points are at
${\pm\qv}^*$ residing in the light tan regions of
Fig.~\ref{fig:phasediagram}(b), and there are symmetry related
regions corresponding to $\Wr_2$ and $\Wr_3$.
When a path through $\Wr_0$ is traversed, beginning on one
edge and ending another, a pair of Weyl points created at ${\rm
M}_j$ pass through the diametrically opposite dark gray regions
of Fig.~\ref{fig:phasediagram}(b) and annihilate at ${\rm
M}_{j\pm 1}$. Similarly, for a path that begins on right
boundary of $\Wr_1$, passes through $\Wr_1$, and
terminates on the left boundary, the Weyl points are born at
$\pm{\rm M}_2$, pass through the opposite tan regions of
Fig.~\ref{fig:phasediagram}(b), and annihilate at $\pm{\rm
M}_3$.       For $x = (-\alpha,
(\alpha+1)/2,(\alpha+1)/2)$ with $\alpha \gg 1$, Weyl points
occur at $\qv^* \approx \pm (2\sqrt{2/\alpha}\,/a)  \hat{x}$,
converging to the $\Gamma$ point as $\alpha\rightarrow\infty$.

Outside the $\Wr_n$ regions there is a gap everywhere in
the Brillouin zone.   There are three disconnected phases that
are topologically distinct.  The topological polarization can
be most easily evaluated at the three simple points (indexed by
$j=1,2,3$) $x_i = \delta_{ij}$.   Then,
\begin{equation}
C(\qv) = e^{i {\qv}\cdot {\dv}_j}.
\label{eq:C100}
\end{equation}
The topological polarization $\Rv_T$ is determined by the
winding numbers $n_i$ of this phase over the independent cycles
of the Brillouin zone:
\begin{equation}
n_i =- {1\over {2\pi i}}\oint_{C_i} C^{-1} dC = -{1\over {2\pi}} \bv_i \cdot \dv_j
\end{equation}
where $C_i$ is the cycle along reciprocal lattice generator
$\bv_i$, which satisfies $\bv_i \cdot \av_j = 2\pi \delta_{ij}$
for Bravais lattice generators $\av_j$. (Note the minus sign in
this equation, which appears because it is defined as an
integral over $ C(\qv)$  rather than $Q(\qv)$ as in
\cite{KaneLub2014}.) Writing
\begin{equation}
\Rv_T = n_1 \av_1 + n_2 \av_2
\end{equation}
then gives
\begin{equation}
\Rv_T = -\dv_j,
\end{equation}
which follows from the ``completeness" relation $\sum_{i=1}^2
\av_i \bv_i = 2\pi I$, where $I$ is the unit matrix. The value
of $\Rv_T$  depends on the real-space positions assigned to the
$A$ and $B$ lattice sites, i.e., on our gauge choice. In our
current symmetric choice, in which the $B$ sites do not sit at
a center of inversion, $\Rv_T$ is not a lattice vector, but the
\emph{differences} between of $\Rv_T$'s in different phases
are. Arrows indicating ${\bf R}_T$ are indicated in
Fig.~\ref{fig:phasediagram}(a). (In the gauge where the origin
lies at an $A$ site, we subtract a particular $\dv_k$ from
each $\dv_j$ in Eq.~(\ref{eq:C100}), and the $n_i$ are all 0 or
$\pm 1$.)

Finally, we note that the line in the phase diagram corresponding
to $x_1=0$ (where the pivot point
is along the bottom edge of the tri-bond) corresponds to a one dimensional limit, in which the system
consists of decoupled horizontal lines that are similar to the SSH model.    In this case there is a direct transition between
topologically distinct gapped phases, which occurs at the blue point $x = (0,1/2,1/2)$.
Here
\begin{equation}
C(\qv) = \cos (q_x a/2) e^{-i q_y a /(2\sqrt{3})},
\end{equation}
so that there is a \emph{line} of zero modes at $q_x = \pm
\pi/a$ along the vertical line that connects ${\rm M}_2$ and
${\rm M}_3$, indicated by a blue line in Fig.~\ref{fig:phasediagram}(b).   This situation is analogous to that in critical
kagome lattices when there are parallel straight lines of bonds
\cite{SunLub2012}. The zero modes are easy to visualize: as we
have discussed, rotation of a  tri-bond about any axis passing
through its pivot point produces no stretch and costs no
energy. Consider a horizontal line of edges containing the
$\dv_2$ and $\dv_3$ vertices of the row of tri-bonds above it
and the $\dv_1$ vertex of the row of tri-bonds below it as
shown in Fig.~\ref{fig:modes}(b).
Rotating neighboring tri-bonds in the upper row by
$\delta\theta$ in opposite directions about the axis passing
through the pivot point and $\dv_1$ (the top vertex) while
rotating neighboring tri-bonds in the lower row in opposite
directions by $\delta\theta/\sqrt{3}$ about the axis along
their bottom edges produces a zero mode. This operation only
affects the given rows, and there is a zero mode for each line
of bonds.

Associated with each zero mode, there must be a state of self
stress. Alternating equal-amplitude stresses on tri-bonds along
any row produces the desired zero-force state. In the case with
$x_1 = 0$ and $x_2 = x_3 = 1/2$, the stress tends to bend the
tri-bonds symmetrically either upward or downward about a line
passing through the pivot point and vertex 1.  Alternation of
the sign of the stresses causes vertices 2 and 3 to experience
equal and opposite forces from the two tri-bonds each shares
along the x axis. When $x_2 \neq x_3$, neighboring rows are
still decoupled, and each is equivalent to the SSH model, whose
critical point occurs when $x_2 = x_3$. Thus the lines defined
by $x_i = 0$ for some $i$, which include the perimeter of the
region where all $x_i$ are positive, correspond to the 1D
limit.

\section{Edge States\label{sec:edge}}
Strips with periodic boundary conditions in one direction or
samples with free sides on all boundaries can be produced by
removing lines of tri-bonds from the system under full periodic
boundary conditions.  Each line of cut tri-bonds creates two
free edges. Since the number of sites and tri-bonds are equal
under periodic boundary conditions, the index theorem reduces
to
\begin{equation}
N_0 - S = \Delta N_B \qquad
\end{equation}
where $\Delta N_B$ is the total number of tri-bonds cut to
produce the free edges. A similar equation applies to each
wavenumber $q$ along the cut producing a strip:
\begin{equation}
n_0 ( q ) - s(q) = \Delta n_B ,
\end{equation}
where $\Delta n_B$ is the number of bonds cut per unit cell of
one of the exposed edges.  How zero modes are distributed on
the free edges depends on the topological polarization $\Rv_T$
and a local surface polarization $\Rv_L$
\cite{KaneLub2014,LubenskySun2015} according to the same
formula derived for central-force Maxwell lattices. The number
of zero modes per edge unit cell (or equivalently per edge 
wavenumber $q$) for a given edge corresponding to a lattice
``plane" indexed by the reciprocal lattice vector $\Gv$
pointing along the edge's \emph{outer normal} is
\begin{equation}
n_0 = (\Rv_T + \Rv_L) \cdot \Gv/(2 \pi ).
\label{Eq:surface_zm2}
\end{equation}
The local polarization $\Rv_L$ is simply the electric
polarization at the given edge that arises from assigning a
charge $+1$ to sites on the $A$-sublattice and a charge $-1$ to
sites on the $B$ sublattice. Of course only components of
$\Rv_L$ parallel to $\Gv$ contribute to $n_0$ so we are free to
add arbitrary components to $\Rv_L$ parallel to the edge .

It is instructive to look at a couple of examples.  Consider
the strip with edges parallel to the $x$-axis as shown in
Fig.~\ref{fig:modes}(b).  To produce this strip, one tri-bond
per surface unit cell had to be removed, so there is a total of
one zero mode per wavenumber on the two exposed edges. The
local polarization $\Rv_L^L$ on the lower edge with outer
surface normal, $\Gv = - 4\pi/{\sqrt{3}a} \hat{y}$, is equally
well represented by $-\dv_1/2$ , $\dv_2$, or $\dv_3$, giving a
local contribution to the number of edge zero modes of $1/3$.
On the upper surface $\Rv_L^U = +\dv_1$, for a contribution of
$2/3$ to edge-mode count.  The topological count for the bottom
surface is respectively $2/3$, $-1/3$, and $-1/3$ for $\Rv_T$
equal to $-\dv_1$, $-\dv_2$, and $-\dv_3$ for a total of one
zero mode on the bottom surface and none on the top surface for
$\Rv_T = -\dv_1$ and no zero mode on the bottom and one the top
surface for $\Rv_T = -\dv_2, -\dv_3$.  A similar analysis for a
strip parallel to the $y$-axis yields for the number of zero
modes on the left and right surfaces $(n_0^R, n_0^L) = (1,1)$,
$(0,2)$, and $(2,0)$ for $\Rv_T$ equal to $-\dv_1$, $-\dv_2$,
and $-\dv_3$, respectively. When there are Weyl points, zero
modes shift from one side of a strip to the other at edge
wavenumbers equal to the projections of the wavenumber of the
Weyl points onto the edge. 

Insight into Eq.~(\ref{Eq:surface_zm2})  can be gained by
considering what happens to $C$ when the sites and tri-bonds
are indexed at different positions (without changing the
lattice itself). Let
\begin{equation}
\Rv_A' = \Rv_A + \Delta\Rv_A \quad {\rm and} \quad \Rv_B' = \Rv_B + \Delta\Rv_B,
\end{equation}
and define the ``gauge-transformed'' compatibility matrix,
\begin{equation}
C'(\Rv_A',\Rv_B') = C(\Rv_A' - \Delta\Rv_A , \Rv_B' - \Delta\Rv_B).
\end{equation}
Then
\begin{align}
C'(\qv) & = e^{-i\qv\cdot(\Delta\Rv_A-\Delta\Rv_B)}C(\qv) \nonumber \\
\ & |C(\qv)|\,e^{-i\qv\cdot(\Delta\Rv_A-\Delta\Rv_B+\Rv_T)}.
\end{align}
Thus if we chose $\Delta\Rv_A - \Delta\Rv_B = \Rv_L$, we find
that the total polarization of $C'(\qv)$ is $\Rv' = \Rv_L +
\Rv_T$ . Let $\qv = (q_{\perp}, q_{\parallel})$, 
where $q_{\perp}$ and $q_{\parallel}$ are, respectively, the  components of $\qv$ perpendicular (positive toward the sample interior) and parallel to the lattice plane in question,
and define $\eta = \exp(iq_{\perp} a_{\perp})$, where $a_{\perp}$ is the depth of the surface unit
cell, and $0 < q_{\perp} \leq |\Gv|$, where $\Gv$ is the inner
normal reciprocal lattice vector associated with the lattice
plane. $C'(\eta, q_{\parallel})$ contains only positive powers
of $\eta$, and thus no poles in $\eta$, and as a result, the
integral
\begin{equation} \label{eq:n0G}
n_0(\Gv) = \frac{1}{2\pi i}\oint d\eta \frac{d}{d\eta} \ln C'(\eta,q_{\parallel}) = \frac{\Gv\cdot (\Rv_L+\Rv_T)}{2\pi}
\end{equation}
counts the number of zero modes at the surface determined by
$\Gv$.  As particular examples, consider the bottom and left
edges in Fig.~\ref{fig:model}. In the first case, $\Rv_L =
-\dv_1 /2$ and
\begin{equation}
C'(\eta,q_x) = x_1\eta+x_2 e^{-iq_x a/2} + x_3 e^{iq_x a/2},
\end{equation}
where $\eta = \exp(iq_y \sqrt{3}\,a/2)$, has at most one zero
and no poles, in agreement with our result that the top and
bottom surfaces can have either one zero mode or none. In the
second case, $\Rv_L = \dv_2$, and
\begin{equation}
C'(\eta,q_x) = x_3 \eta^2 + x_1 \eta + x_2 e^{iq_y \sqrt{3}\, a/2},
\end{equation}
where $\eta = \exp(iq_x a/2)$.  Again, there are no poles, but
the highest power of ${\eta}$ is 2, and according to
Eq.~(\ref{eq:n0G}), there can be 0, 1, or 2 zero modes in
agreement with our previous results.

When there are Weyl modes, the number of zero modes on a free edge of a strip will change when $q_{||}$ passes through the projection of a Weyl point onto that edge~\cite{BurkovBal2011a,RocklinLub2016}.  The total number of zero modes does not change at this transition, so there must be a change of the opposite sign in the number of zero modes on the opposite edge. In other words, zero modes move from one side of the sample to the opposite at a projection of a Weyl point.  A similar phenomenon occurs at domain walls in systems under periodic boundary conditions, in which the number of zero modes equals the number of states of self-stress and is given by 
\begin{equation}
\nu_T = \Gv\cdot (\Rv_T^1 - \Rv_T^2)/2 \pi ,
\end{equation}
where $\nu_T$ is equal to the number of zero modes per
wavenumber if it is positive and minus the number of states of
self stress if it is negative.  Thus a change in the number of zero modes on a zero-mode domain wall, which occurs at $q_{\parallel}$ equal to the projected Weyl wavenumber, must be accompanied by an equal change in the number of Weyl states of self-stress on a self-stress domain wall.

\section{Physical Models}
\label{sec:phys-models}

\begin{figure}
\includegraphics[width=0.55\columnwidth]{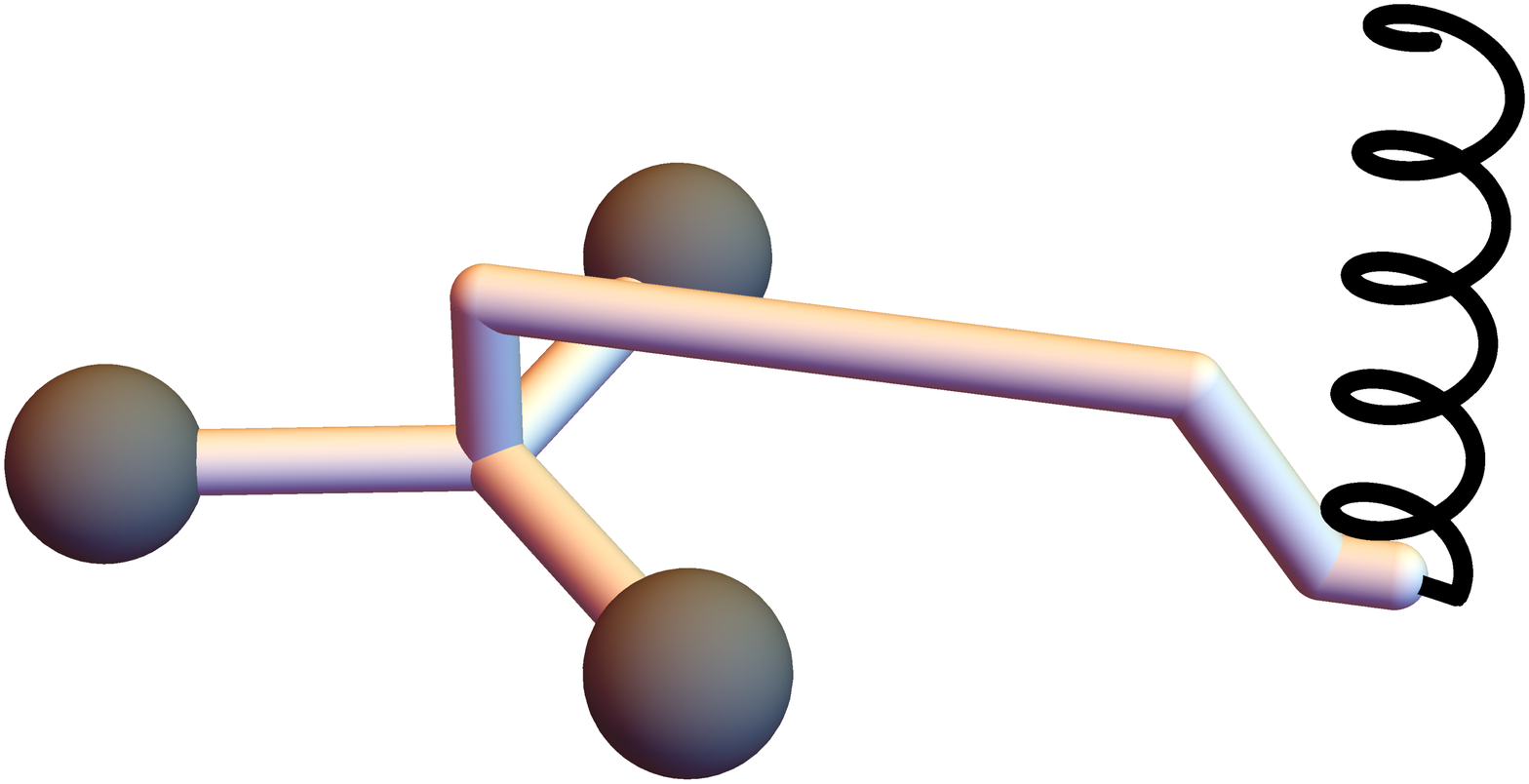}
\includegraphics[width=0.4\columnwidth]{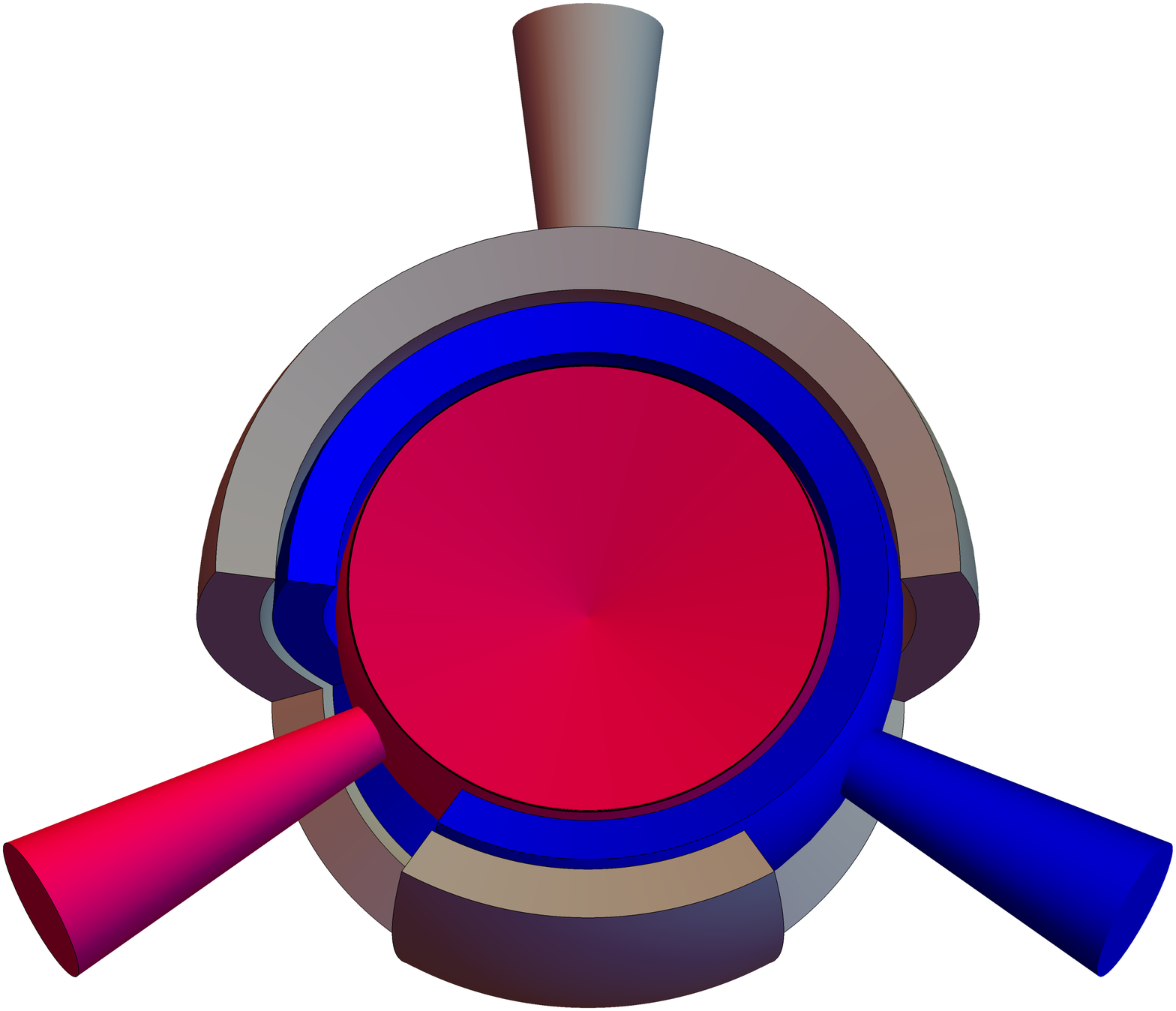}
\caption{Left: Tri-bond ``plate'' with pivot at an arbitrary location.  The bars are assumed to be rigid and have negligible mass compared to the balls.  Right: Cutaway view of a three-pronged ball joint for connecting three plates to a single mass.  The holes in the outer two spherical shells allow for rotations and twists of the each of the three arms about any axis through the common center of the spheres.}
\label{fig:balljoint}
\end{figure}

Physical realization of the mechanical graphene model poses some technical challenges, but
some straightforward approaches are possible.
The simplest version of the tri-bond uses a spring that
directly measures the stretch.  The tri-bond consists of a
rigid plate suspended at its pivot point on a vertical spring
that is attached to a rigid ceiling.  The plates are connected
at each vertex to a ball of mass $m$ through a mechanism that
allows free rotations about the axes passing through the center
of the ball and the pivot point on the plate.  A possible
design with three nested universal ball joints is shown in
Fig.~\ref{fig:balljoint}.
By attaching a lever arm to each plate that extends arbitrarily
far from its centroid and attaching the spring to the end of it
as shown in Fig.~\ref{fig:balljoint}, the pivot point can be
placed at any location in the plane, thus allowing the
realization of the complete phase diagram of
Fig.~\ref{fig:phasediagram}.

The height of the pivot point is precisely $\sum x_i z_i$, so
the difference in energy in the spring (and gravity) from the
equilibrium configuration in which all plates are horizontal is
exactly the desired tri-bond energy.  To mimic the spectrum of
graphene, we need the kinetic energy matrix expressed in terms
of the corner height variables $z(\Rv_A)$ to be a multiple of
the identity: $E_K = (1/2) m \sum_A \dot{z}(\Rv_A)^2$. Thus we
take the mass of a plate to be negligible compared to the mass
of a corner ball. Note that the deviation from coplanarity of
the plate corners and pivot point, which defines the tri-bond
stretch, is measured with respect to the fixed equilibrium
position of the pivot point rather than the pivot point that
moves with the rigid plate (i.e., the position where the spring
is attached).

In the limit of small deviations from equilibrium, this model
directly mimics the generic discussion of the tri-bond above.
Because the tri-bond stretch is directly encoded in a single
spring and the height variables are literally encoded as
heights of the ball joints, both $\ev$ and $\zv$ are
immediately visible.  For example, the self-stress state of the
one-dimensional row of Fig.~\ref{fig:modes} is simply an
alternating pattern of tension and compression in the springs
along that row.

Other physical models could exhibit the mechanical
graphene phase diagram, if not the exact same spectrum.  
All that is required is a potential
energy of the form
\begin{equation}
U = \sum_B \frac{1}{2} k(x_1,x_2,x_3) (x_1 z_1 + x_2 z_2 + x_3 z_3)^2,
\end{equation}
where the effective stiffness $k$ may depend on the location of
the pivot.  Because the features of greatest interest are zero
modes, the form of the kinetic energy does not affect their
existence or locations in the Brillouin zone, though it will
affect the details of the spectrum at finite frequency if the
kinetic energy is not simply a multiple of
$\dot{z}_1^2+\dot{z}_2^2+\dot{z}_3^2$.

A natural example realizes the tri-bond energy as the bending
energy of a triangular elastic plate that is pinned at the
pivot point and attached at each vertex to two other plates.
While the precise form of the lowest potential energy of the
sheet for arbitrary (small) choices of the corner heights is
difficult to calculate due to the boundary conditions at the corners and on
the free edges, it must vanish for all rigid rotations of the
plate, for which $x_1 z_1 + x_2 z_2 + x_3 z_3 = 0$, and thus
cannot depend on any other linear combination of the corner
heights.

Another possibility consists of rigid plates coupled pairwise
by springs at their corners.  In
this model, there are two degrees of freedom per plate (the two
rocking angles) and thus two vibrational bands.  One band is fully
gapped, consisting of modes in which the net displacement of
the three plate corners at any given $A$ site is zero.  The
other band exhibits the mechanical graphene phases, with each
plate effectively acting as a tri-bond that couples the average
displacements at each $A$ site. 

\section{Concluding remarks}

This paper has introduced a mechanical model that is a precise analog
of the tight-binding model of graphene, and defines an appropriate two dimensional
generalization of the SSH analog introduced in Ref.~\cite{KaneLub2014}.     This model system exhibits a rich phase
diagram of Weyl phases, along with gapped phases with distinct topological polarizations.
Our proposed structures are amenable to physical implementation, and it will be interesting
to construct them and to probe their mechanical mode structures.

In addition, our construction introduces the tri-bond, which opens a new avenue for studies
of Maxwell lattices.  It will be interesting to use this approach, and generalizations of it, to construct
new classes of two and three dimensional mechanical systems.

\begin{acknowledgments}
The authors are grateful for the hospitality of the Lorentz
Institute of for Theoretical Physics at the University of
Leiden, where work on this project began. This work was
supported in part by NSF grant DMR-1104707 (TCL), by two
Simons Investigator grants (TCL and CLK).
\end{acknowledgments}

\bibliographystyle{unsrt}
\normalbaselines
\bibliography{./RPP}

\end{document}